\documentclass[1p]{elsarticle}
\usepackage{epsfig}
\usepackage{color}           
\usepackage{epsfig}

\begin{document}

\begin{frontmatter}

\title{Influence of competition in minimal systems with discontinuous absorbing
phase transitions}

\author[if-ufrgs,df-ufpr]{Salete Pianegonda}

\author[if-usp]{Carlos E. Fiore\corref{carlos}}
\ead{fiore@if.usp.br}

\cortext[carlos]{Corresponding author}

\address[if-ufrgs]{Instituto de F\'{\i}sica, Universidade Federal 
do Rio Grande do Sul, Caixa Postal 15051, CEP 91501-970, 
Porto Alegre, RS, Brazil}

\address[df-ufpr]{Departamento de F\'{\i}sica, Universidade Federal 
do Paran{\'a}, Caixa Postal 19044, CEP 81531-980, 
Curitiba, PR, Brazil}

\address[if-usp]{Instituto de F\'{\i}sica,
Universidade de S\~{a}o Paulo, 
Caixa Postal 66318\\       
05315-970 S\~{a}o Paulo, SP, Brazil}
%\date{\today}

\begin{abstract}
Contact processes (CP's) with particle creation requiring a minimal
neighborhood (restrictive or threshold CP's) present
a novel sort of discontinuous absorbing transitions, 
that revealed itself robust under the inclusion of different ingredients, such as 
distinct lattice topologies, particle annihilations and diffusion. Here, we 
tackle on the influence of competition between restrictive
and standard dynamics (that describes the usual CP and a continuous 
DP transition is presented). Systems
have been studied via mean-field theory  (MFT)
and numerical simulations.
Results show partial contrast between MFT and numerical results. While
the former predicts that considerable competition rates are required
to shift the phase transition, the latter reveals the change
occurs for rather limited (small) fractions. Thus,  unlike
previous ingredients (such as diffusion and others),
limited competitive rates suppress the phase coexistence.

\end{abstract}

%\maketitle

\end{frontmatter}

%\linenumbers

\section{Introduction}

The usual contact process \cite{r4} is probably the simplest
example of system presenting  an absorbing phase transition.  It is
composed of two subprocesses:   ``spontaneous''
annihilation and ``catalytic'' particle creation, in which 
new species are created only in empty sites on the neighborhood
of at least one particle.  
Despite the lack of an exact solution, its phase
transition and critical behavior are very well known and belong 
to the robust directed percolation (DP) universality class 
\cite{henkel,r1,r2}.
Many  generalizations  of its
rules can be extended
not only for theoretical purposes, but also
for the description of a large variety of  systems in the framework
of physics \cite{fiore2,dickman,ginelli}, chemistry \cite{zgb,evans}, 
ecology \cite{pnas} and others.
In these cases, the 
competition among dynamics leads to several
new findings. In some cases \cite{tome,fiore2,dickman}, 
 the competition between particle hoping (diffusion)
and  annihilation of three adjacent particles (instead
of a single particle) is responsible for a reentrant
phase diagram and a stable active phase  for extremely low activation rates.
In other examples \cite{salete}, by allowing particles to have different
creation rates with respect to their first and second neighbors, 
the competition is responsible for the appearance of an
active asymmetric phase with spontaneous breaking symmetry.
Also, when
particles interact in a symbiotic manner \cite{scp,scp2},  two
active symmetric phases, in which only one species is present, emerges.

An interesting generalization  are the called 
restrictive (threshold) CPs, in which the phase transition changes
from continuous to discontinuous for $d \ge 2$. They are similar
to the usual CP, but 
%Recently, it was also shown that a class of contact processes, 
%differing slightly from their respective
%original models seems to be the simplest one-species lattice model
%yielding a discontinuous transition for $d \ge 2$.
one  requires at least
two particles for creating a new species (in the original CP 
at least one particle is needed). Different restrictive models
have revealed that the  phase transition remains
first-order by including   other sorts of creation 
\cite{evans,evans2,oliveira2,oliveira,carlos},
annihilation rules \cite{carlos}
and also  different lattice topologies  \cite{durret}. 
Besides, a very recent study \cite{salete2} has claimed that the particle
diffusion does not suppress the phase coexistence, in partial
contrast with results obtained from stochastic differential
equation approaches \cite{pnas}.

In order to enhance the understanding about the robustness of discontinuous
absorbing phase transitions presented in such simple (minimal) models,
here we  extend the study undertaken in Ref. \cite{salete2} by addressing 
the competition between the above restrictive
and standard dynamics (that describes the usual CP). 
In the last case, the transition 
belongs to the usual and robust directed percolation (DP) universality class.
 More specifically, the particle creation requiring at least 
one and two nearest neighbor 
particles is chosen with distinct (but complementary) 
probabilities. Two distinct restrictive rules will be considered.
%The extreme  cases reduce to the restrictive models
% and the usual CP \cite{sabag}, respectively.
Although the phase transition is expected to be continuous (discontinuous)
in the extreme regimes of less (more) frequent restrictive interactions 
\cite{oliveira2,oliveira}, 
our study focus on  answering the following questions:
(i) is the phase coexistence suppressed for the inclusion 
of limited fraction of non-restrictive
interactions? (ii) or on the contrary, only large rates are required 
for shifting the phase transition?
(iii) Finally, how do the difference of models
influence  the phase transition and corresponding tricritical points (separating phase
coexistence from critical transition)? 

Models will be investigated in the framework of mean field theory (MFT)
and numerical approaches, which lead (as will be shown further) to conclusions
in partial agreement. 
While the MFT predicts the 
inclusion of considerable fraction
of non-restrictive interactions is needed for shifting  the phase transition, 
numerical simulations 
show the suppression occurs for limited small rates.
A second contribution concerns the establishment of precise
approaches for characterizing the discontinuous transitions. 
Although the critical exponents for DP phase transitions are well known, 
no established scaling behavior is
known for the discontinuous case. As it will be shown, 
our methodology clearly distinguishes continuous from discontinuous
transitions, reinforcing previous claims about the existence of a common
finite-size scaling for the latter case \cite{carlos,salete2,fioremarc}.

This paper is organized as follows: In Sec. II we define the models and
mean-field predictions are presented in Sec. III. Sec. IV 
shows the numerical results and  conclusions are drawn 
in Sec. V.

\section{Models}

Systems are defined on a square lattice of size $L$ and 
each site has an occupation variable $\sigma_i$ 
that assumes the value $0$ or $1$ depending whether  it
is empty or occupied, respectively. 
The dynamics is composed of the following ingredients: particle annihilation
and creation requiring a minimal neighborhood $nn$ of at least $1$ and
$2$ particles. More precisely, particles 
are annihilated with rate $\alpha$ and with probabilities $p$ 
and $1-p$ the particle creation requires at least
$nn \ge 1$ and $nn \ge 2$ adjacent particles, respectively.  
Here, we consider two different rules for  the second
creation subprocess.   
In the first case (rule A), the  creation occurs
 with rate $nn/4$ 
\cite{oliveira2}, whereas in the latter (rule B) the creation is
always $1$ (provided $nn \ge 2$ in both cases) \cite{oliveira}.
Thus, while for the rule A  particles are created with rate 
proportional to the
number of their nearest neighbors, it is independent of $nn$
for the rule B.

The extreme cases $p=0$ and $p=1$ reduce themselves to the restrictive 
models investigated in Refs. \cite{oliveira2,oliveira} 
(rules A and B) and the 
usual CP, respectively. The
transitions are discontinuous and continuous yielding
$\alpha_{0}=0.2007(6)$ (rule A and $p=0$) \cite{oliveira2}, 
$\alpha_{0}=0.352(1)$ (rule B and $p=0$) \cite{oliveira}
and $\alpha_c=0.60653(1)$ \cite{sabag} for $p=1$.
In all cases, the  order parameter is the particle density $\rho$, in such a
way that $\rho=0$ in the absorbing state and $\rho \neq 0$ 
in the active phases.

\section{Mean-field analysis}
Since all above models present no exact solution,
the first inspection over the effect of competition can be
 undertaken by performing mean-field analysis (MFT). 
Starting from the master equation, we derive 
relations for appropriate quantities 
and truncate the associated probabilities.
In the first level of approximation (one-site mean-field),
the generic probability $P(\sigma_0,\sigma_1,...,\sigma_{n-1})$
is rewritten as a product of one-site probabilities, in such
a way we have
${P(\sigma_0,\sigma_1,...,\sigma_{n-1})=P(\sigma_0)P(\sigma_1)...P(\sigma_{n-1})}$. Due to the 
relation $\sum_{\sigma_{i}=0}^1 P(\sigma_i)=1$
only one relation is sufficient for the analysis.

In order to obtain
improved results, we include  correlation
of two sites. This can be done by performing the pair mean-field approximation 
that consists of rewriting
the $n$-site probabilities ($n>2$) as
products of two-site probabilities yielding
\begin{equation}
P(\sigma_0,\sigma_1,...,\sigma_{n-1}) \simeq \frac{P(\sigma_0,\sigma_1) 
P(\sigma_0,\sigma_2)...P(\sigma_0,\sigma_{n-1})}{P(\sigma_0)^{n-2}}\,.
\label{pair_approx}
\end{equation}
In this case, two equations are required to calculate
the system properties. By identifying the system density $\rho$ as the one-site probability 
$\rho=P(1)$ and considering
the two-site correlation
$u=P(01)$, from the previous
models' rules we obtain the following relations
\begin{eqnarray}
\frac{d\rho}{dt}&=&(1-p)[2P(01100)+P(01010)+3P(01011)\nonumber\\ 
&&+P(01111)]+pP(01)
-\alpha P(1),
\label{ra}
\end{eqnarray}

\begin{eqnarray}\label{ra2}
\frac{du}{dt}&=&(1-p)[-\frac{3}{2}P(01011)-P(01111)]\\
%\\ \nonumber
%\label{ra2}
&&+\frac{p}{2}[-3P(011)+P(01)]-\alpha\,P(01)+
\alpha\,P(11),
\nonumber
%\label{ra2}
\end{eqnarray}

for the rule A  and
\begin{eqnarray}
\frac{d\rho}{dt}&=&(1-p)[4P(01100)+2P(01010)+4P(01011)\nonumber\\
&&+P(01111)]
+p\,P(01)-\alpha P(1),
\label{rb}
\end{eqnarray}

\begin{eqnarray}\label{rb2}
\frac{du}{dt}&=&(1-p)[-2P(01011)-P(01111)]\\
&&+\frac{p}{2}[-3P(011)+P(01)] -\alpha P(01)+\alpha P(11),
\nonumber
\end{eqnarray}
for the rule B.
The symbol $P(\sigma_0,\sigma_1,\sigma_2,\sigma_3,\sigma_4)$ 
 denotes the probability
of finding the central site in the state $\sigma_0$ and its four nearest
neighbors in the states $\sigma_1$, $\sigma_2$, $\sigma_3$ and $\sigma_4$.

From the one-site MFT,  Eqs. (\ref{ra}) and (\ref{rb}) read
\begin{equation}
\frac{d \rho}{dt}=(1-p)\rho^2(1-\rho)[3-3\rho+\rho^2]+p \rho(1-\rho)-
\alpha \rho,
\label{mfta}
\end{equation}
for the rule A and 
\begin{equation}
\frac{d \rho}{dt}=(1-p)\rho^2(1-\rho)[6-8\rho+3\rho^2]+p \rho(1-\rho)-
\alpha \rho,
\label{mftb}
\end{equation}
for the rule B. The stationary values and the order
of phase transition  are obtained by taking $\frac{d\rho}{dt}=0$.
At the level of two-sites correlations Eqs. 
(\ref{ra}) and (\ref{ra2}) read
\begin{equation}
\frac{d\rho}{dt}=(1-p)\left[\frac{3u^{2}}{(1-\rho)}-\frac{3u^{3}}{(1-\rho)^{2}}+
\frac{u^{4}}{(1-\rho)^{3}}
\right] +pu-\alpha\rho,
\end{equation}
\begin{eqnarray}
  \frac{du}{dt}&=&(1-p)\left[-\frac{3u^{3}}{2(1-\rho)^{2}}+\frac{u^{4}}{2(1-\rho)^{3}}
\right]\nonumber  \\ 
&&+p\left[\frac{u}{2}-\frac{3u^{2}}{2(1-\rho)}\right]-2\alpha\,u+\alpha\rho,
\end{eqnarray}
for the rule A and Eqs. (\ref{rb}) and (\ref{rb2}) read
\begin{equation}
\frac{d\rho}{dt}=(1-p)\left[\frac{6u^{2}}{(1-\rho)}-\frac{8u^{3}}{(1-\rho)^{2}}+
\frac{3u^{4}}{(1-\rho)^{3}}\right]+p\,u-\alpha\rho,
\end{equation}
\begin{eqnarray}
\frac{du}{dt}&=&(1-p)\left[-\frac{2u^{3}}{(1-\rho)^{2}}+\frac{u^{4}}{(1-\rho)^{3}}
\right]\nonumber\\
&&+p\left[\frac{u}{2}-\frac{3u^{2}}{2(1-\rho)}\right]-2\alpha\,u+\alpha\rho, 
\end{eqnarray}
for the rule B.
In this approximation, the steady solutions are obtained 
by taking $\frac{d\rho}{dt}=\frac{du}{dt}=0$, from which
we locate
the transition point and  the order 
of transition by solving a system of two coupled equations 
for a given set of parameters ($\alpha$,\,$p$).  In similarity
to the one-site MFT, the classification
of phase transition is determined by inspecting the behavior
of $\rho$ vs $\alpha$. The existence of a spinodal
behavior (with $\rho$ increasing by raising $\alpha$) signals a 
discontinuous transition, whereas its absence
is consistent with a continuous transition. 
In practice, we considered distinct $\alpha$'s differing for a fixed increment
$\delta \alpha$   ($\alpha_n=\alpha_{n-1}+\delta\alpha$, 
where $\delta\alpha=0.005$) and 
a spinodal behavior can be verified in a very clear way. 
Since no procedures similar to the ``Maxwell construction'' 
are available for the nonequilibrium case, the coexistence points 
will be estimated by the maximum value  of $\alpha$
with the spinodal behavior replaced by a jump in $\rho$. 
In particular, the tricritical point has been estimated as the point from which the spinodal behavior is not discernible (within the numerical precision).  
Fig. 1 $(a)$ and $(b)$ show the phase diagram 
 for distinct weakening parameters $p$ of the restrictive creation.
In particular, both single and pair MFT predict that  considerable 
fractions $p_{t}$ are required to suppress the phase coexistence 
for both rules A and B. In the former approximation, 
the crossover of regimes occurs for  $p_t=0.74(1)$ (rule A) 
and $0.85(1)$ (rule B), whereas somewhat lower 
 values $p_{t}=0.67(1)$ (rule A) and 
 $p_t=0.80(1)$ (rule B) are required for the latter one.
Thus, MFT results predict that large competition rates are 
required to suppress the phase coexistence.
 However, both levels of approximation show that, in contrast
to the rule A, the phase diagram
exhibits a reentrant shape for the rule B for a limited range
of $p$. Also, in the two cases,
%Unlike previous results,  inset shows that for $p=0.9$ 
%$\rho$ vanishes continuously 
%at $\alpha_c=0.675$, consistent to a second-order transition.
 along the critical
line  $p$ and $\alpha$ present a dependence about linear.
%------------------------------                                                
\begin{figure}[h]
\centering
\includegraphics[scale=0.37]{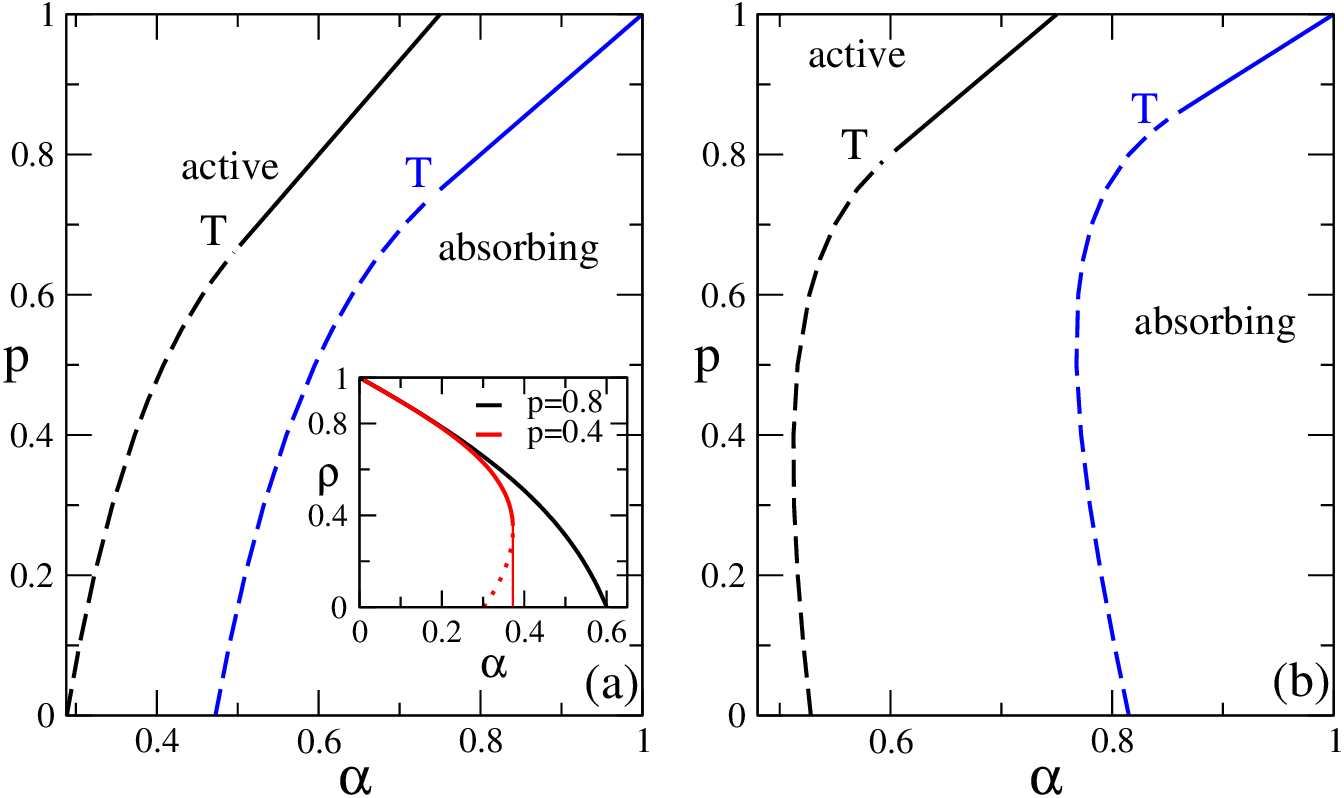}
\caption{({\bf color online:}) MFT phase diagrams for  the rules A $(a)$ and
B $(b)$ at the level of one (right curves) and two sites (left curves). 
Dashed and continuous lines
denote discontinuous and continuous phase transitions,
respectively. The symbols $T$ denote the tricritical points.
For the rule A, inset shows  the density $\rho$ vs $\alpha$ 
for $p=0.8$ and $p=0.4$. In the latter case, the spinodal (dotted line)
has been replaced by a jump in $\rho$.}
\label{fig11}
\end{figure}
%---------------------------------      

\section{Numerical results}
In order to draw conclusions beyond the mean-field
level, we perform numerical analysis for both models A and B. 
 Although  the overestimation  of the tricritical points under the MFT 
is expected, since  approximated methods typically predict  superior limits
than "exact" ones, we investigate if numerical simulations also
predict limited or large (non-restrictive)  interactions 
for the suppression of the phase coexistence.
Numerical simulations have been carried out for distinct system sizes 
and periodic boundary conditions.
Although the behavior of
continuous transitions is well established,  this is not
the case of discontinuous ones. For this reason, 
in the first analysis we study the time decay of  the density $\rho$
starting from a fully occupied lattice as initial configuration. 
As for critical 
and discontinuous phase transitions, for small
$\alpha$ the density $\rho$ converges to a
 definite value indicating endless activity,
in which particles are continuously 
created and annihilated. On the contrary, for 
larger $\alpha$'s one expects an exponential
decay of $\rho$ toward a complete particle extinction.
At the critical point $\alpha_c$, the density $\rho$ behaves algebraically 
following a power-law behavior $\rho \sim t^{-\theta}$,  being 
$\theta$ its associated critical exponent. 
For the DP universality class $\theta$ 
is well known and reads $\theta=0.4505(10)$ \cite{henkel}.
Conversely, some recent papers \cite{carlos,munoz14} 
have proposed that the  discontinuous transition point 
$\alpha_{0}$ can be estimated as 
the separatrix between active and absorbing regimes. In such cases,
the absence of a power law behavior separating the indefinite activity
from the absorbing states is verified and it will be used as the first 
evidence of a discontinuous
phase transition.

The second approach for classifying the phase transition
is obtained by performing
numerical simulations in the steady regime.  
For the continuous case, the ratio  
$U_{2}=\frac{\langle \rho^2 \rangle}{\langle \rho \rangle^{2}}$  
is an useful  quantity, since its
evaluation for distinct system's sizes 
cross at the critical point $\alpha_c$, 
with a well defined  value $U_{2c}$. 
Also, at   $\alpha_c$ the system density $\rho$ and its variance 
$\chi=L^{2}[\langle \rho^{2}\rangle-\langle \rho \rangle^{2}]$
follow the scaling behaviors
$\rho \sim L^{-\beta/\nu_{\perp}}$ and $\chi \sim L^{\gamma/\nu_{\perp}}$,
being $\beta/\nu_{\perp}$ and $\gamma/\nu_{\perp}$
their associated critical exponents.  Moreover, at the vicinity
of the critical point, one expects that $\rho$ vanishes
algebraically following the relation $\rho \sim (\alpha_c-\alpha)^\beta$. 
For the DP universality class, all above quantities present
precise values given by $U_{2c}=1.3257(5)$, $\beta/\nu_{\perp}=0.796(9)$,
$\gamma/\nu_{\perp}=0.41(2)$ and $\beta=0.5834(30)$ \cite{henkel}.
In contrast,  discontinuous transitions are signed by
bimodal probability distribution $P_\rho$ (characterizing
the absorbing and active phases) and also by a peak in the variance 
$\chi$.
For equilibrium systems, the maximum of $\chi$ and other quantities scale
with the system volume and its position $\alpha_L$
obeys the asymptotic relation $\alpha_L =\alpha_0 - c/L^2$ \cite{rBoKo,rBoKo2},
being $\alpha_0$ the transition point in the
thermodynamic limit and $c$ a constant. 
Recent papers \cite{carlos,sinha,salete2,fioremarc}
have shown that  similar scaling is verified for nonequilibrium
phase transitions.
Alternatively, the transition point
can also be estimated as
the value of $\alpha_L$ in which the two
peaks of the probability distribution have equal weights (area) \cite{sinha,fioremarc}.

In order to obtain the steady quantities, we apply the models dynamics together 
with the quasi-steady method \cite{qs}.
It consists of  storing a list of $M$ active configurations (here we
store $M=2000-3000$ configurations)
and whenever the system falls into the absorbing state a
configuration is  randomly extracted from the list.
The ensemble of stored configurations is continuously updated, where
for each MC step a configuration belonging to the list is replaced with
probability ${\tilde p}$ (typically one takes ${\tilde p}=0.01$)
by the actual system configuration, provided it is not absorbing.

\begin{figure}[h]
\centering
\includegraphics[scale=0.343]{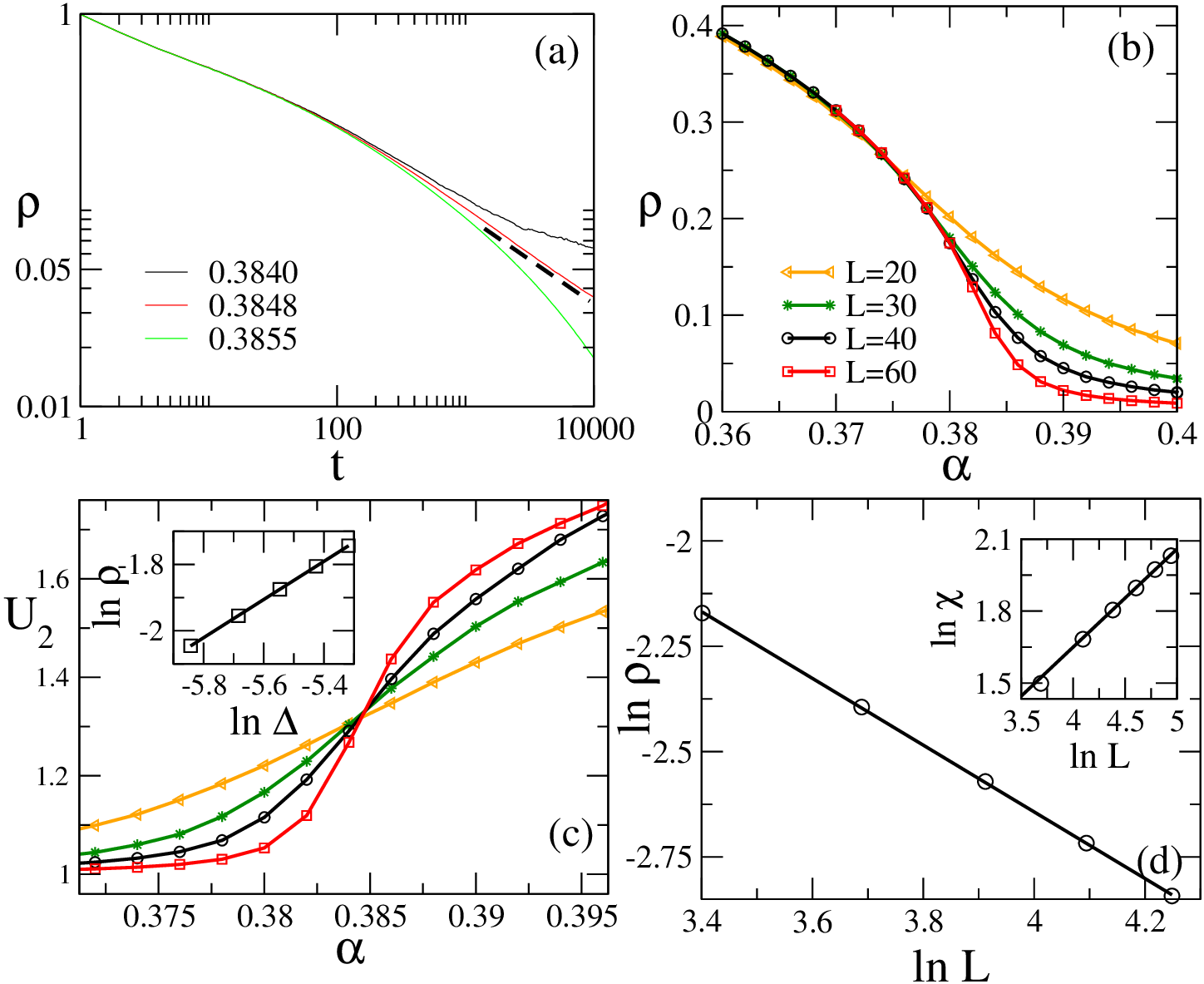}
\caption{({\bf color online:}) For the rule A and $p=0.5$, panel $(a)$
 shows the time evolution of $\rho$ for distinct $\alpha$'s. The
straight line has slope $0.4505(10)$.
Panel $(b)$ shows $\rho$ vs $\alpha$ 
for distinct system sizes $L$.  
In $(c)$  we show   $U_2$ vs $\alpha$ and         
inset shows a log-log plot of $\rho$ vs $\Delta=\alpha_c-\alpha$.
The straight line has slope  $\beta=0.58(1)$.
In $(d)$ the log-log plot of $\rho$ and $\chi$ (inset)
vs $L$ at the critical point $\alpha_c=0.3848(3)$.    
The slopes read $\beta/\nu_{\perp}=0.79(1)$
and  $\gamma/\nu_{\perp}=0.41(2)$ (inset).  }
\label{fig3}
\end{figure}

\begin{figure}[h]
\centering
\includegraphics[scale=0.343]{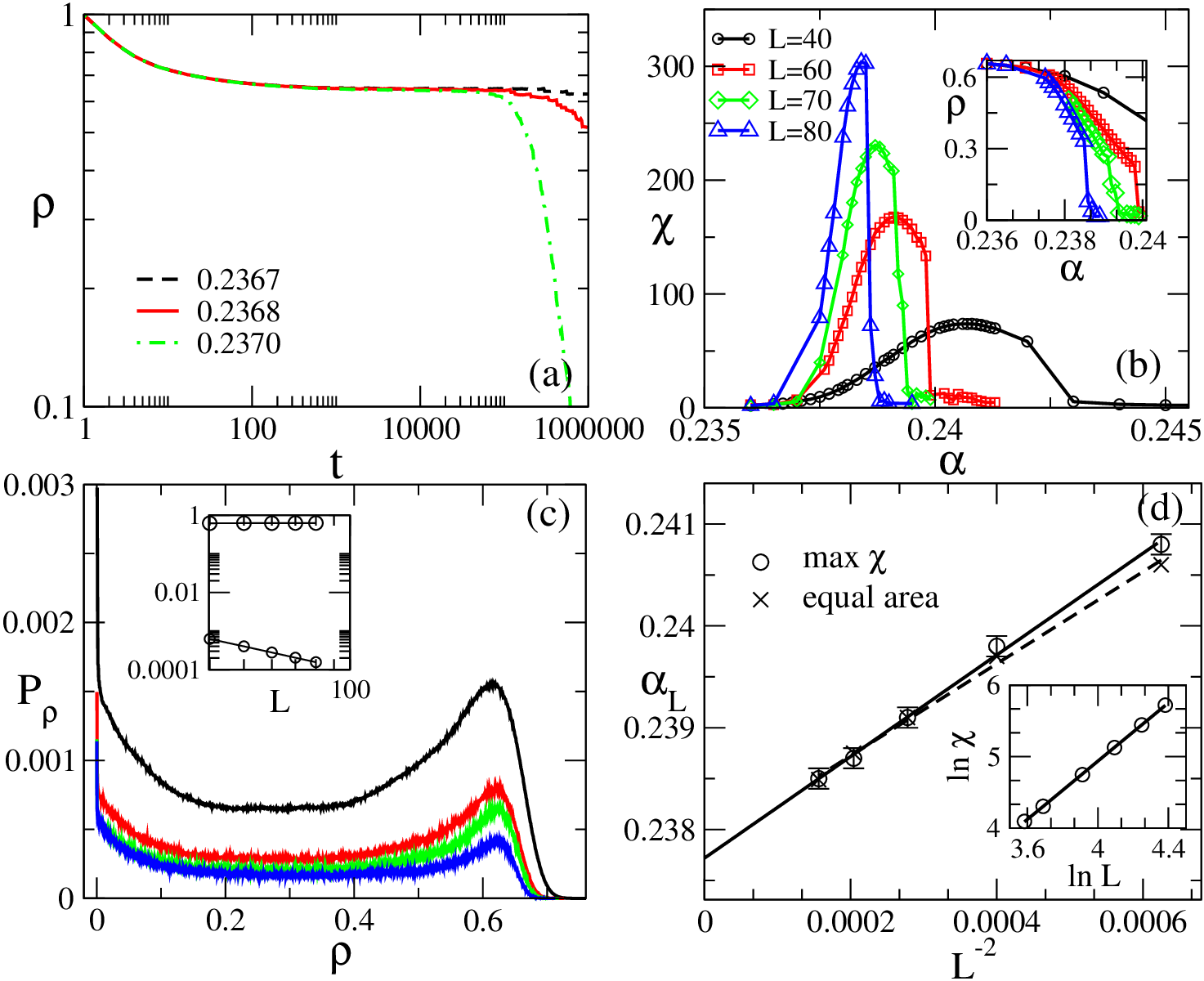}
\caption{({\bf color online:}) For the rule A and $p=0.1$, panel $(a)$ shows 
the time evolution of $\rho$ for distinct $\alpha$'s. In $(b)$
$\chi$ and $\rho$ (inset) vs $\alpha$ 
for distinct system sizes $L$. In $(c)$ the 
quasi-stationary probability distribution                        
$P_{\rho}$ having peaks                                       
with the same area for different $L's$. In the inset, the  log-log plot                          
of the quasi-steady densities vs L.
In $(d)$, the scaling plot        
of $\alpha_{L}$, in which $\chi$ is maximum and peaks of $P_{\rho}$ have
equal area, vs $L^{-2}$. Inset shows the      
log-log plot of the maximum of $\chi$ vs $L$ and the straight line has         
slope $2$.                                                                    }
\label{fig2}
\end{figure}

For the rule A, results are summarized in Figs. \ref{fig3} and \ref{fig2}
for two representative values  $p=0.5$ and $p=0.1$, respectively.
%------------------------------
In the former case,  the time evolution of $\rho$ 
(panel $(a)$) shows that a power-law decay  with exponent  consistent  
with the DP value $0.4505(10)$ is verified at $\alpha_c \sim 0.3848$, 
consistent with  a continuous transition.  Panels $(b)$ and $(c)$ reinforce
this finding. In particular,
the vanishing of $\rho$ is followed by a crossing of
  the ratio $U_2$ evaluated for different $L's$, in which all curves 
intersect  at    $\alpha_c=0.3848(3)$ with $U_{2c}=1.32(2)$, 
also in consistency with the DP value $1.3257(5)$.
Moreover, at the critical point,  the log-log plots of
$\rho$ vs ${\rm L}$ (panel (d)) and $\rho$ vs $\Delta$
(inset of panel (c) where $\Delta \equiv \alpha_c-\alpha$) 
provide the  critical exponents $\beta/\nu_{\perp}=0.79(1)$  
and $\beta=0.58(1)$. They are in a good agreement with the DP
values $0.796(9)$ and $0.5834(30)$, respectively.  
Finally, analysis of the variance $\chi$  (inset of panel $(d)$) also provides an exponent
consistent to the DP value  $\gamma/\nu_{\perp}=0.41(2)$  \cite{henkel}.

%---------------------------------

An entirely distinct behavior is verified for $p=0.1$ (Fig. \ref{fig2}). 
There is a  threshold value $\alpha_0 \sim 0.2368$ separating active from 
the exponential decay of $\rho$  (panel $(a)$).
Also,  bimodal probability distributions  with
 active $\rho_{ac}$ and absorbing $\rho_{ab}$ 
densities yielding distinct dependencies on $L$ 
(panel $(c)$ and its inset), reinforce a discontinuous
transitions for such case. 
Third, $\rho$ vanishes in a tiny 
interval of $\alpha$ followed by peaks
of $\chi$ (panel $(b)$ and its inset), whose  positions  
scale with $L^{-2}$ (panel (d)),
from which we obtain the estimate $\alpha_0=0.2377(2)$.
Similar scaling is also verified for the positions $\alpha_L$'s 
in which the peaks have equal area (panels $(c)$ and $(d)$), 
providing the value 
$\alpha_0=0.2378(2)$ that is very close to the above one. 
Inset of panel (d) shows the log-log plot of the maximum of
$\chi$ vs the system's size $L$ (the straight line has slope 2).

\begin{figure}[h]
\centering
\includegraphics[scale=0.35]{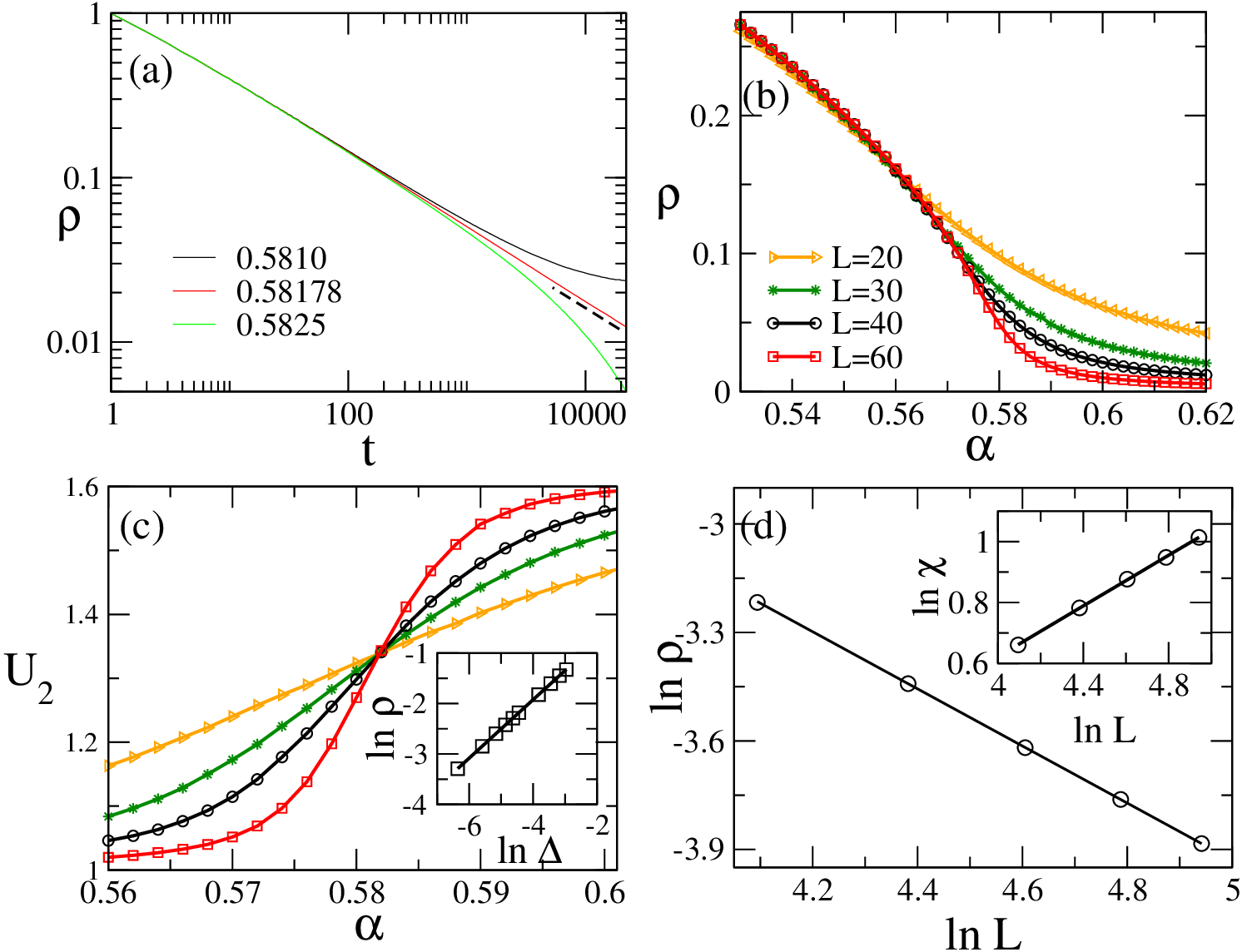}
\caption{({\bf color online:}) For the rule B and $p=0.8$, 
 panel $(a)$
 shows the time evolution of $\rho$ for distinct $\alpha$'s. The
straight line has slope $0.4505(10)$.
Panel $(b)$ shows $\rho$ vs $\alpha$ 
for distinct system sizes $L$.  
In $(c)$  we show   $U_2$ vs $\alpha$ and         
inset shows a log-log plot of $\rho$ vs $\Delta=\alpha_c-\alpha$.
The straight line has slope  $\beta=0.58(1)$.
In $(d)$ the log-log plot of $\rho$ and $\chi$ (inset)
vs $L$ at the critical point $\alpha_c=0.5818(3)$.    
The slopes read $\beta/\nu_{\perp}=0.79(1)$
and  $\gamma/\nu_{\perp}=0.41(2)$ (inset).}
\label{fig5}
\end{figure}

\begin{figure}[h]
\centering
\includegraphics[scale=0.35]{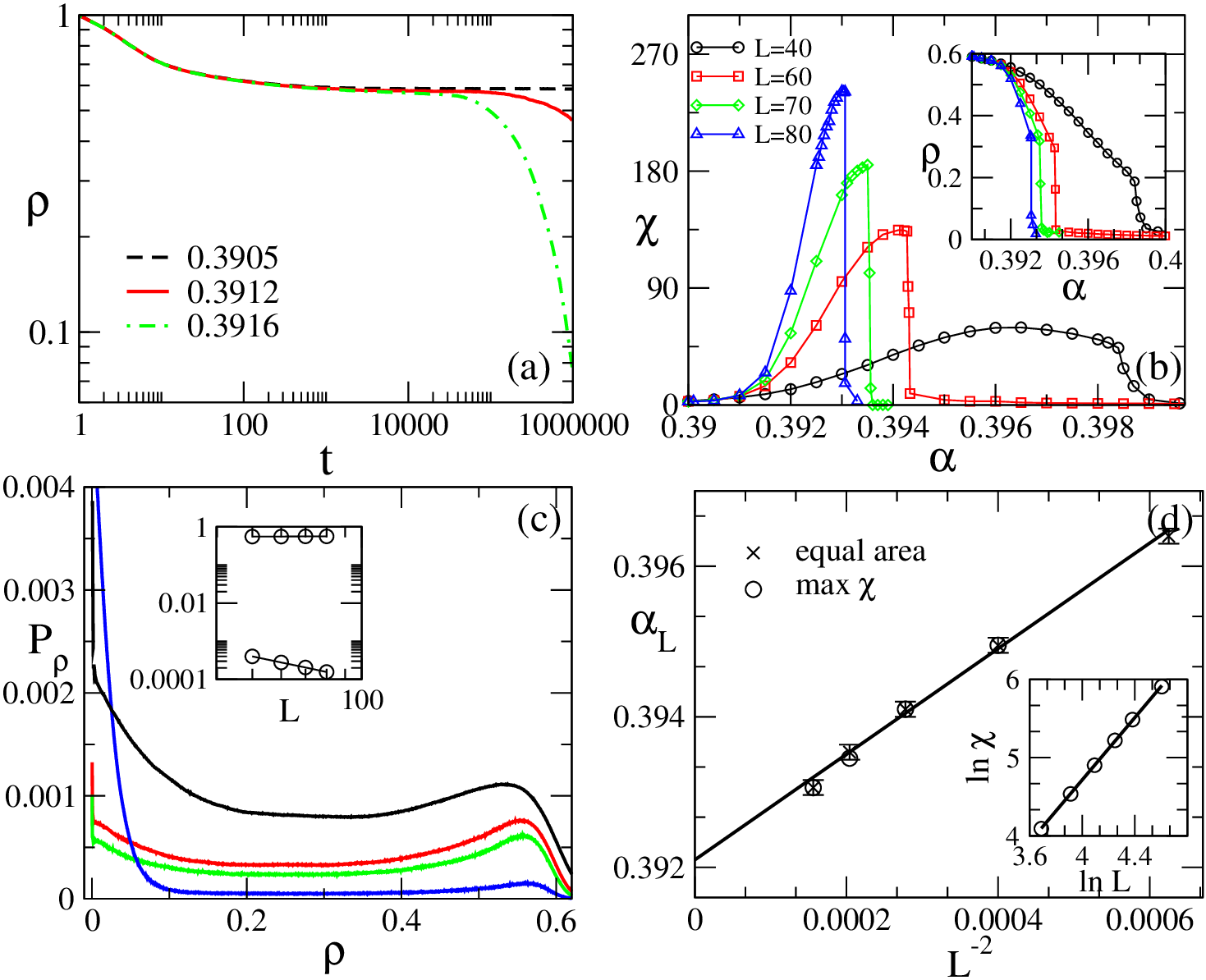}
\caption{({\bf color online:}) For the rule B and $p=0.06$,  panel $(a)$ shows 
the time evolution of $\rho$ for distinct $\alpha$'s. In $(b)$
$\chi$ and $\rho$ (inset) vs $\alpha$ 
for distinct system sizes $L$. In $(c)$ the 
quasi-stationary probability distribution                        
$P_{\rho}$ having peaks with the same area for different  $L$'s.
In the inset, the  log-log plot                          
of the quasi-steady densities vs L.
In $(d)$, the scaling plots        
of $\alpha_L$, in which $\chi$ is maximum
and peaks of $P_{\rho}$ have equal area, 
vs $L^{-2}$. Inset shows the      
log-log plot of the maximum of $\chi$ vs $L$ and the straight line has         
slope $2$.                                              }
\label{fig4}
\end{figure}

%---------------------------------
In Figs. \ref{fig5} and \ref{fig4}, we extend the analysis for the rule B.
As for the rule A, the behavior for larger values of $p$ 
is consistent with a critical transition, from which we obtain for $p$=0.8 (Fig. \ref{fig5})
the values $\alpha_c=0.58178(5)$ (time decay of $\rho$ in panel (a))
and $\alpha_c=0.5818(3)$ (crossing of the $U_2$ curves in panel (c)), 
in excellent agreement to each other.
Also, all critical exponents are consistent with the previous DP
 values (inset of panel $(c)$).
Conversely,  low   $p$'s shows that 
the phase transition is signed by the absence of a
power-law behavior and the
probability distributions presenting
bimodal shapes, consistent with a discontinuous transition.
For example, for $p=0.06$ (Fig. \ref{fig4}) we obtain the estimate $\alpha_0 \sim 0.3912$ (from time decay of $\rho$),
which is close to the values
$\alpha_0=0.3918(4)$ (maximum of $\chi$) and $0.3920(4)$ 
(equal area  of $P_{\rho}$).

\begin{figure}[h]
\centering
\includegraphics[scale=0.37]{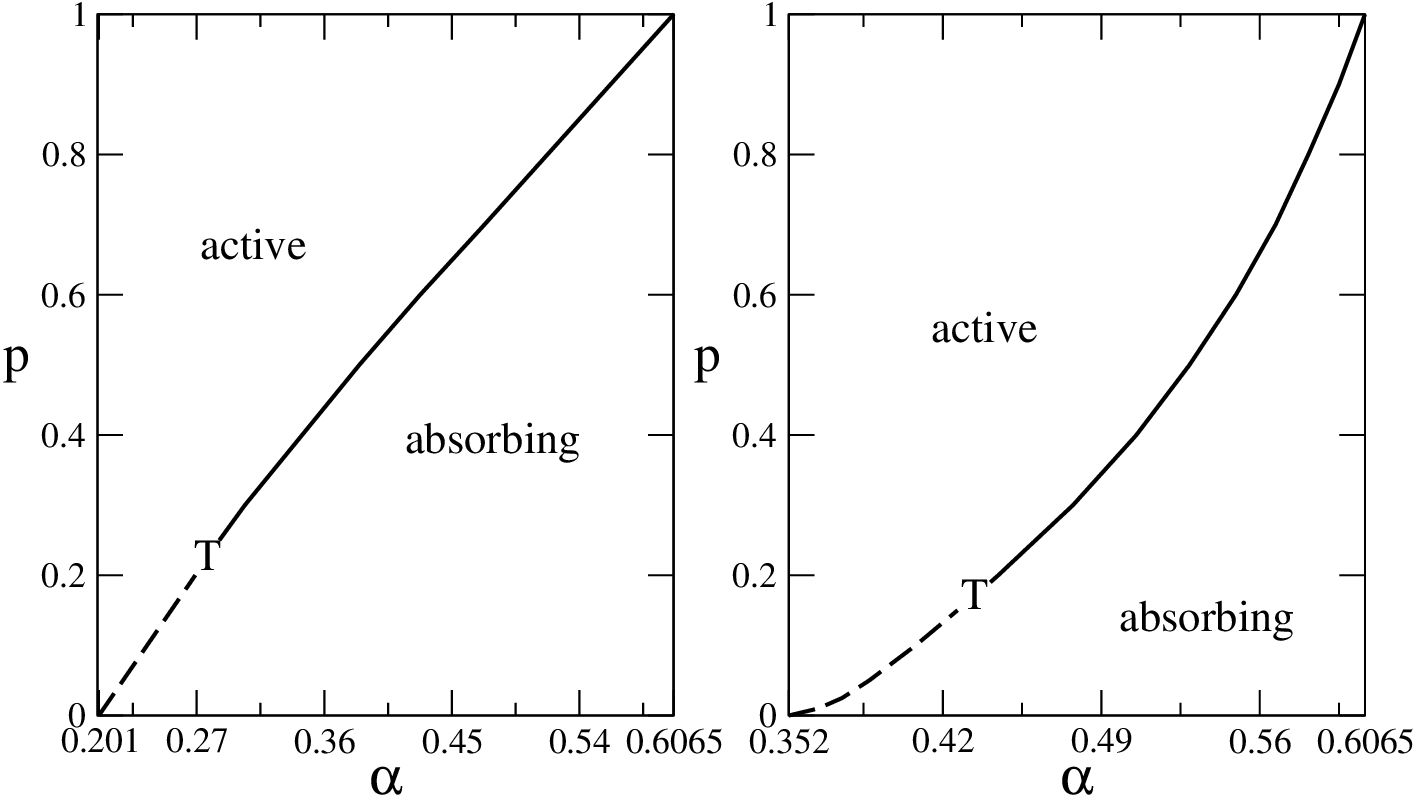}
\caption{The phase diagrams in the plane $p-\alpha$ obtained from MC simulations for the 
rules A (left) and B (right). Dashed and continuous lines
denote discontinuous and continuous phase transitions,
respectively,  and $T$ is the tricritical point.}
\label{fig8}
\end{figure}

Extending above analysis for distinct values of $p$ we obtain the 
phase diagrams for rules A and B shown
in Fig. \ref{fig8}. 
In both cases, continuous and discontinuous transition lines
are separated by tricritical points $T$'s 
located  at $(\alpha_t,p_t)=(0.276(9),0.25(4))$ 
and $(\alpha_t,p_t)=(0.435(10),0.15(4))$, for the rules A 
and B, respectively.  The uncertainties have been estimated
taking into account that  close to $T$, there
is a crossover regime signed by  exponents
continuously varying with $p$.
Thus, in contrast with MFT, numerical results  reveal that
limited (instead of large)  values of $p_t$ separate phase coexistence
from the criticality. Other differences concern that
although the tricritical value of $p_t$ of both models are close to each
other, the coexistence line of the rule A is somewhat larger than the one for rule.
Thus, the specific model rules have few influence in the crossover 
between criticality and phase coexistence.
As a difference between numerical results
and MFT, phase diagrams  are not reentrant.  
A last comment concerns  that both tricritical
points $p_t$'s  are quite larger than the  value $p_t \sim 0.032$
obtained  from the model from Ref. \cite{evans3}.
 Unlike the present studied models, such  
restrictive version   requires at least two 
adjacent diagonal pairs of particles for creating new species.
Consequently,  unlike
our results, the phase transition is characterized by a generic two-phase
coexistence and exhibits an interface orientational dependence at the transition
 point.  Thus, the difference among models can be responsible for
prolonging the discontinuous transition line in our cases.
%---------------------------------
\section{Conclusions}
Aimed at giving a further step for addressing the robustness
of discontinuous absorbing phase transitions under the presence of distinct ingredients,
we consider the competition between
dynamics in two minimal models. They have been investigated under mean-field
approaches and two sorts of numerical simulations.
Results  showed that limited competition rates are sufficient for 
 the suppression of phase coexistence. This is in partial
contrast with previous results \cite{carlos,salete2}
in which the inclusion of ingredients like  particle 
diffusion and distinct annihilation
rules do not shift the discontinuous transitions.
As a final comment, we mention that 
all results for the first-order transitions reinforce previous
claims over a common finite-size scaling
for nonequilibrium transitions \cite{carlos,salete2,fioremarc}, in which
in similarity with the equilibrium case,
 relevant quantities scale with the system volume.

\section*{Acknowledgments}
The authors wish to thank Brazilian scientific agencies 
CNPq, FAPESP, INCT-FCx for the financial
support. Salete Pianegonda also wishes to thank the Physics Department of the Federal Technological
University, Paran\'{a} (DAFIS-CT-UTFPR) for providing the access to its high-performance computer facility.

\section*{References}

\bibliography{mybibfile}

\end{document}